\documentclass{article}

\usepackage{color}
\usepackage{graphicx}
\usepackage[labelfont=bf,textfont={sl,bf},lofdepth,lotdepth]{subfig}
\usepackage{gensymb}
\usepackage{amssymb}

\author{Denis Terwagne$^\dagger$ and John W.M. Bush$^\ddagger$\\
\\\vspace{6pt} $^\dagger$GRASP, D\'epartement de Physique, \\\vspace{6pt} 
Universit\'e de Li\`ege, B-4000 Li\`ege, Belgium\\\vspace{6pt} 
$^\ddagger$Department of Mathematics,\\\vspace{6pt} 
Massachusetts Institute of Technology, 02139 Cambridge, USA } 
\title{Tibetan Singing Bowls}

\begin{document}
\maketitle
%\tableofcontents

\section*{Abstract}

We present the results of an experimental investigation of the acoustics and 
fluid dynamics of Tibetan singing bowls. Their acoustic behavior is rationalized in terms
of the related dynamics of standing bells and wine glasses.
% and our study allows us to infer the Young's modulus. 
Striking or rubbing a fluid-filled bowl excites wall vibrations, and concomitant
waves at the fluid surface. Acoustic excitation of the bowl's natural vibrational modes
allows for a controlled study in which the evolution of the surface waves with increasing
forcing amplitude is detailed. Particular attention is given to rationalizing the observed
criteria for the onset of edge-induced Faraday waves and droplet generation via surface 
fracture. Our study indicates that drops may be levitated on the fluid surface, induced to 
bounce on or skip across the vibrating fluid surface.

\section{Introduction}

\begin{figure}[h]
\begin{center}
\includegraphics[width=10cm]{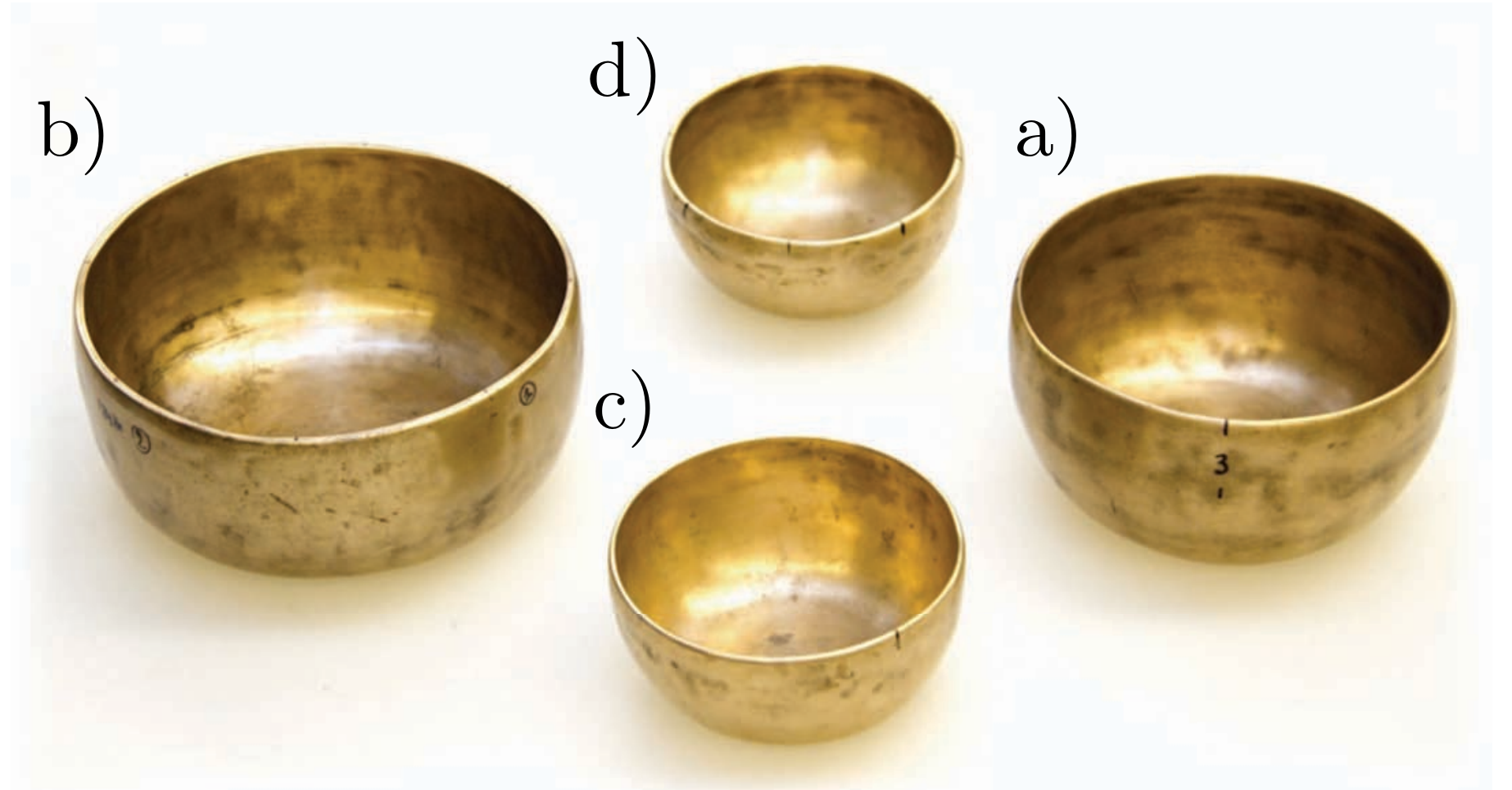}
\end{center}
\caption{Our Tibetan singing bowls: a) Tibet 1, b) Tibet 2, c) Tibet 3 and d) Tibet 4. \label{figbowl}}
\end{figure}

Tibetan singing bowls are thought to have originated from Himalayan fire cults of the 5th century BC and  have since been used in various religious ceremonies, including shamanic journeying and meditation.  The Tibetan singing bowl (see Fig.~\ref{figbowl}) is a type of standing bell played by striking or rubbing its rim with a wooden or leather-wrapped mallet. This excitation causes the sides and rim of the bowl to vibrate and produces a rich sound. Tibetan bowls are hand made and their precise composition is unknown, but generally they are made of a bronze alloy that can include copper, tin, zinc, iron, silver, gold and nickel. When the bowl is filled with water, excitation can cause ripples on the water surface. More vigorous forcing generates progressively more complex surface wave patterns and ultimately the creation of droplets via wave breaking. We here quantify this evolution, and demonstrate the means by which the Tibetan singing bowl can levitate droplets. 

Related phenomena are known to appear in other vessels, including the Chinese singing bowl, on which vibrations are generated by rubbing the vessel's handles with moistened hands. The more familiar vibration of a wine glass is produced by rubbing its rim with a moist finger. The dependence of the glass's vibration frequency on its material properties, geometry, and characteristics of the contained fluid was elucidated by French \cite{French1982} and subsequent investigators \cite{Chen2005A, Chen2005B, Jundt2006, Courtois2008}. The coupling between two singing wine glasses has been investigated by Arane {\it et al.} \cite{Arane2009}. Apfel \cite{Apfel1985} demonstrated experimentally that wine glass vibration generates capillary waves near the walls of a fluid-filled glass. Moreover, he made the connection between these waves and the acoustic whispering gallery modes elucidated by Rayleigh \cite{Rayleigh1945}. By studying the deformation and the sound spectrum produced by a single wine glass, Rossing \cite{Rossing1990} elucidated the mechanism of the glass harmonica \cite{Rossing1993}, an instrument designed by Benjamin Franklin. Joubert \textit{et al.} \cite{Joubert:2007} provided a theoretical rationale for observations of standing waves in a singing wine glass.
The Tibetan singing bowl has to date received relatively little attention. Inacio \emph{et al.} examined experimentally the acoustic response of bowls excited by impact and rubbing \cite{Inacio2006}, and
proposed a dynamical formulation of the bowl and presented some numerical simulations \cite{Inacio2004}. 
The hydrodynamics of a fluid-filled Tibetan bowl will be the focus of our investigation.

In 1831, Faraday \cite{Faraday1831} demonstrated that when a horizontal 
fluid layer is vibrated vertically, its interface 
remains flat until a critical acceleration is exceeded. Above this threshold, a field 
of waves appears at the interface,  parametric standing waves oscillating with half the forcing frequency \cite{Rayleigh:1883, Benjamin1954, Miles1990, Kumar1994, Edwards1994, Kudrolli1996}. The form of such Faraday waves depends on the container geometry; however boundary effects can be minimized by using a large container. 
The Faraday waves have a wavelength prescribed by the relative importance of surface tension and gravity,
being capillary and gravity waves in the short and long wavelength limits, respectively. 
As the forcing acceleration is increased, progressively more complex  wave patterns arise, and the 
interfacial dynamics become chaotic \cite{Kudrolli1996, Chen:1997, Edwards1994}. Ultimately, large
amplitude forcing leads to surface fracture and the ejection of droplets from the fluid bath.
%Breaking of Faraday waves when bowls or wine glasses are excited vigorously, leads to droplet ejection from the edge towards the center. 
A recent study on the breaking of Faraday waves in a vertically shaken bath has been performed in both the capillary and gravity wave limits by Puthenveettil \textit{et al.} \cite{Puthenveettil2009}. Goodridge \textit{et al.} \cite{Goodridge1997} studied the drop ejection threshold of capillary waves in a glycerine-water solution for frequencies up to 100~Hz.

%When a Tibetan singing bowl contains liquid, waves reminiscent of those in the wineglass may be observed. 
Faraday \cite{Faraday1831} reported that such parametric waves can also be emitted by a vertical plate plunged into a liquid bath and oscillating horizontally: along both sides of the plate, waves aligned perpendicular to the plate oscillate at half the forcing frequency. These so-called cross-waves, or edge-induced Faraday
waves, are typically produced by a wave-maker, and have received considerable attention \cite{Garrett:1970, Barnard:1972, Mahony:1972, Miles:1988a, Becker:1991}. Hsieh \cite{Hsieh2000} examined theoretically wave generation in a vibrating circular elastic vessel, specifically the axisymmetric capillary waves and circumferential crispations that appear in an inviscid fluid subject to radial wall displacement. These studies have demonstrated that the excitation of the cross waves is due to a parametric resonance. The complexity of this problem lies in the nonlinear interactions  between the motion of the oscillating rim and the resulting wave field.
%amplitude of the waves launched on the surface.
% These edge-induced Faraday waves are similar to those observed in a vertically shaken bath \cite{Benjamin1954}. This latter problem can be solved by considering that the liquid bath is subject to an oscillating gravity. 

Droplets ejected on the liquid surface by breaking Faraday waves may bounce, skid and roll before coalescing. A number of recent studies have examined droplets bouncing on a vertically vibrating liquid bath below the Faraday threshold \cite{Couder2005, Gilet2008, Dorbolo2008}.  The air film between the drop and the liquid surface is squeezed and regenerated at each successive bounce, its sustenance precluding coalescence and enabling droplet levitation. A similar effect arises on a soap film, a system more readily characterized theoretically \cite{Gilet:2009}. The bouncing periodicity depends on the size of the drop and the vertical forcing acceleration of the bath \cite{protiere:2005,eddi:2008}. Couder \textit{et al.} \cite{Couder2005} have shown that, through the waves emitted at previous bounces, some droplets can walk horizontally across the liquid surface. Several factors are needed to sustain a so-called ``walker'' \cite{Eddi2011}. First, the drop must bounce at half the forcing frequency, so that it resonates with the resulting Faraday wave field. Second, the bath must be close to the Faraday instability threshold so that Faraday waves of large amplitude and spatial extent can be excited by the drop impacts. The droplet bounces on the slope of the wave emitted at the previous bounce and so receives an impulsive force in a specific direction, along which it walks with a constant speed. Such walkers have both wave and particle components, and have been shown to exhibit quantum-like dynamical behaviour previously thought to be peculiar to the microscopic realm \cite{couder:2006, eddi:2009a, fort:2010, bush:2010}. Might such modern physics
arise in our ancient bowls?

The manuscript is divided into two main parts. In \S 2, we examine the acoustics of the Tibetan singing bowls. 
We begin in \S 2.1 by reviewing the related dynamics and theoretical description of the wine glass \cite{French1982}. In \S 2.2, our bowls are presented and their deformation spectra analyzed. Then, by 
adapting the theoretical description of the vibrating wine glass, we infer the Young's modulus of the alloy comprising our bowls. 
In \S 3, we examine the dynamics of flows generated within liquid-filled vibrating bowls. A review of Faraday waves and droplet ejection on a vertically shaken bath is presented in \S 3.1. In \S 3.2, our experimental technique is detailed. In \S 3.3, we analyze the surface waves generated on the liquid bath, and their relation
to Faraday waves. In \S 3.4, we examine the limit of large amplitude forcing, in which droplets are ejected 
by breaking Faraday waves. Comparisons are made with experiments performed on a vertically shaken liquid bath. Droplet levitation is considered in \S 3.5, where particular attention is given to developing criteria
for droplet bouncing and exploring the possibility of walking droplets. A summary of our results
is presented in \S 4.

\section{Acoustics} \label{SecAcoustics}

\subsection{Background} \label{SecBackground}

\begin{figure}[h]
\begin{center}
\includegraphics[width=12cm]{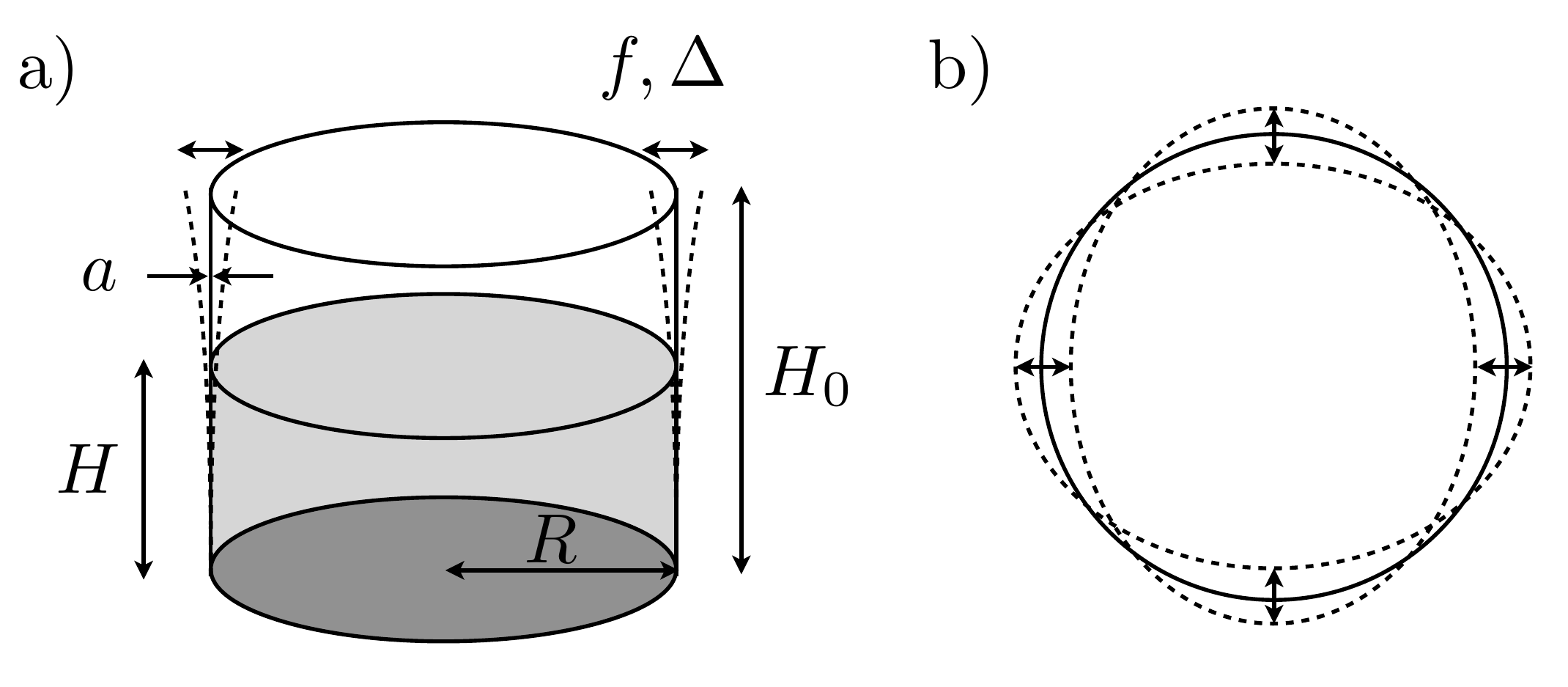}
\end{center}
\caption{a) Schematic illustration of a vibrating vessel filled with liquid. Relevant parameters are the height $H_0$ and the radius $R$ of the vessel, the thickness of the rim $a$, the liquid level $H$, the frequency $f$ and the amplitude $\Delta$ of the oscillating rim. b) Top view of a vibrating vessel in its fundamental mode (2,0), characterized by its 4 nodes and 4 antinodes. \label{Expschema}}
\end{figure}

Both the wine glass and the Tibetan bowl can be excited by either tapping or rubbing its rim. We denote by ($n$,$m$) the vibrational mode with $n$ complete nodal meridians and $m$ nodal parallels. Tapping excites a number of vibrational modes \cite{Rossing1993}, while rubbing excites primarily the (2,0) fundamental mode (see Fig.~\ref{Expschema}b). Entirely flexural motion implies radial and tangential displacements proportional to $n\sin{n\theta}$ and $\cos{n\theta}$, respectively, $\theta$ being the azimuthal coordinate \cite{Perrin1985}. For the (2,0) mode the maximum tangential motion is necessarily half the maximum normal motion.

A leather mallet can excite bowl vibrations via a stick-slip process, as does a finger moving on a wineglass. The moving mallet forces the rim to follow the mallet during the stick phase; during the slip phase, the bowl rim relaxes to its equilibrium position. This rubbing results in a sound composed of a fundamental frequency plus a number of harmonics. While the mallet is in contact with the bowl, one of the nodes follows the point of contact \cite{Joubert:2007}, imparting  angular momentum to the bound liquid.

To simplify the acoustic analysis, one can approximate the glass or bowl by a cylindrical shell with a rigid base and an open top (Fig.~\ref{Expschema}a). The system can then be described in terms of 7 physical variables, the radius $R$, height $H_0$, thickness $a$, Young's modulus $Y$ and density $\rho_s$ of the cylindrical shell, and the frequency $f$ and amplitude $\Delta$ of its oscillating rim. The system can thus be described in terms of 4 independent dimensionless groups, which we take to be $R/H$, $\Delta/a$, $\Delta/R$ and a Cauchy number $Ca=\rho_s f^2 \Delta^2 / Y$ that indicates the relative magnitudes of the inertial and the elastic forces experienced by the vibrating rim.

The sound is emitted by bending waves that deform the rim transversely as they propagate. The speed $V_b$ of bending waves on a two-dimensional plate of thickness $a$ is given by \cite{Kinsler1982}
\begin{equation}
V_b=\left(\frac{\pi V_L fa}{\sqrt{3}}\right)^{1/2},
\end{equation}
where $V_L$ is the longitudinal wave speed in the solid (approximately $5200\,\rm{m/s}$ in glass). In order for the bending wave to traverse the perimeter in an integer multiple of the period, we require
\begin{equation}
\frac{1}{f} \propto \frac{2\pi R}{V_b}.
\end{equation}
Thus, since $V_b \sim \sqrt{f a}$, we have $f \propto a/R^2$: the frequency increases with rim thickness, but decreases with radius.

A more complete theoretical analysis of the wine glass acoustics  \cite{French1982} can be applied to our Tibetan bowls. An ideal cylinder fixed at the bottom is considered (Fig.~\ref{Expschema}a), its wall vibrating with largest amplitude at its free edge or rim. The rim's kinetic energy and elastic energy of bending in the mode (2,0) are calculated by assuming that the radial position is proportional to $\cos{2\theta}$, with $\theta$ being the azimuthal coordinate. By considering conservation of total energy (kinetic plus elastic bending energy), an expression for the frequency of this mode can be deduced:

\begin{equation}
f_0=\frac{1}{2\pi}\left(\frac{3Y}{5\rho_s}\right)^{1/2}\frac{a}{R^2}\left[1+\frac{4}{3}\left(\frac{R}{H_0}\right)^{4}\right]^{1/2}.
\end{equation}
When the bowl is partially filled with liquid to a depth $H$ (Fig.~\ref{Expschema}a), the frequency decreases. French \cite{French1982} captured this effect by considering the kinetic energy of the liquid near the rim, and so deduced the frequency of the fundamental mode:
\begin{equation}
\left(\frac{f_0}{f_H}\right)^2\sim 1+ \frac{\alpha}{5}\frac{\rho_l R}{\rho_s a} \left(\frac{H}{H_0}\right)^4, \label{f0surfH}
\end{equation}
where $\rho_l$ is the liquid density and $\alpha \sim 1.25$ is a constant indicating the coupling efficiency between the rim and fluid displacements. Similarly, frequencies of higher modes can be calculated by considering a radial position proportional to $\cos{n \theta}$ and with $m$ nodal parallels \cite{French1982}:

\begin{equation}
f_{(n,m)}=\frac{1}{12\pi}\left(\frac{3Y}{\rho_s}\right)^{1/2}\frac{a}{R^2}\left[\frac{(n^2-1)^2+(mR/H_0)^4}{1+1/n^2}\right]^{1/2} \label{fnm}
\end{equation}

\subsection{Tibetan bowls} \label{SecTibetanBowls}

Four different antique bowls of different sizes have been studied (Fig. 1). They are referred to as Tibet 1, 2, 3 and 4 and their physical characteristics are reported in Table 1.

\begin{table}[h]
\begin{center}

\begin{tabular}{|c|c|c|c|c|}
   \hline
   Bowl name &  Tibet 1 &  Tibet 2 &  Tibet 3 &  Tibet 4\\
   \hline
   Symbol & $\bullet$ & $\circ$  & $\boxtimes$  & $\blacklozenge$ \\
   $f_{(2,0)}$ (empty) (Hz) & 236 & 187 & 347 & 428 \\
   Radius $R$ (cm) & 7.5 & 8.9 & 6 & 5.9 \\
   Thickness $a$ (cm) & 0.34 & 0.38 & 0.31 &0.37 \\
   Mass (kg) & 0.690 & 0.814 & 0.306 &0.312 \\
   Volume (cm$^3$) & 76& 97& 35 & 33\\
   Density (kg/m$^3$) & 9079 & 8366 & 8754 & 9372 \\
   \hline
\end{tabular}
\label{TableTibet}
\end{center}
\caption{Physical properties of the four Tibetan bowls used in our study.}
\end{table}

\begin{figure}[h!]
\begin{center}
\includegraphics[width=12cm]{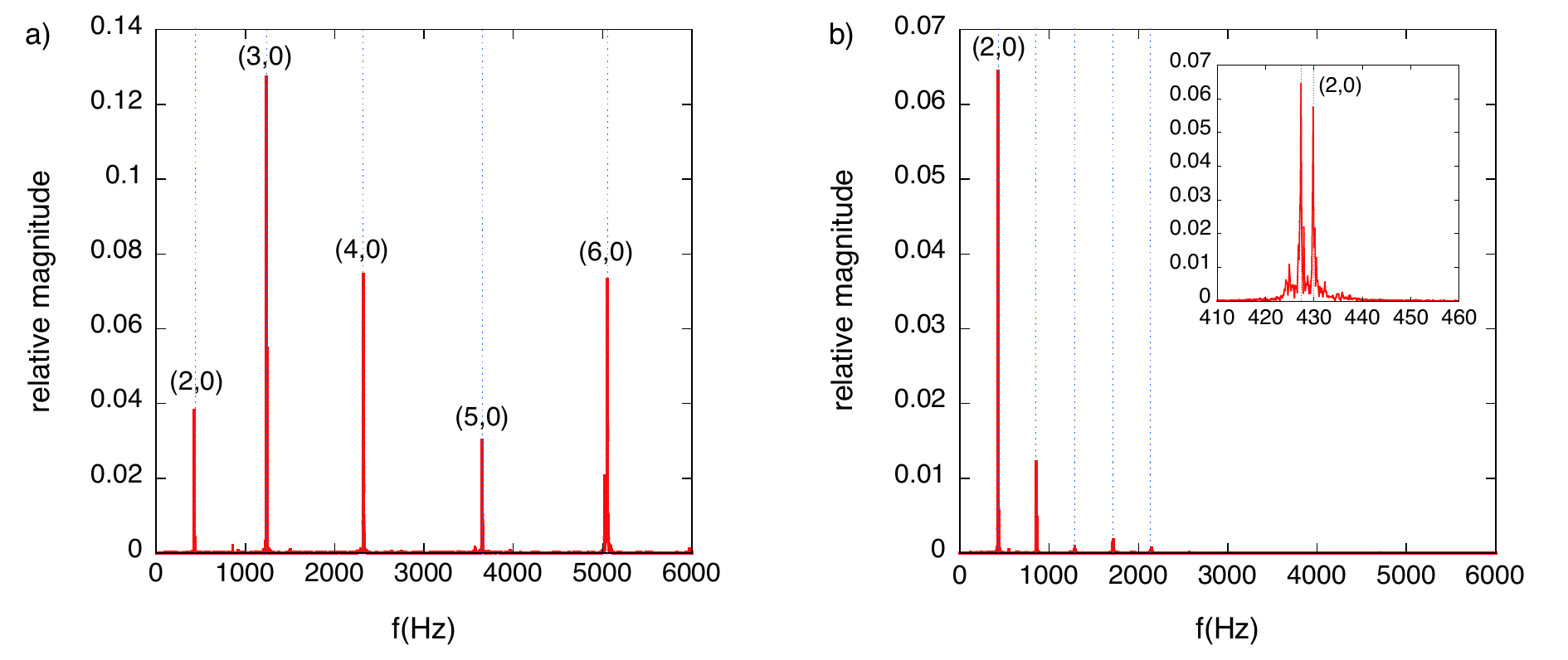}
\end{center}
   \caption{a) Frequencies excited in the bowl Tibet 4 when struck with a wooden mallet. The different peaks correspond to the natural resonant frequencies of the bowl, and the associated deformation modes $(n,m)$ are indicated. b)  Excited frequencies of the bowl Tibet 4 when rubbed with a leather mallet. The first peak corresponds to the deformation mode (2,0) and subsequent peaks to the harmonics induced by mode lock-in. In the inset, a magnification of the first peak provides evidence of its splitting due to the asymmetry of the bowl.}
  \label{spectraglass}
\end{figure}

When a bowl is struck or rubbed, the sound emitted by the resulting bowl vibrations is recorded with a microphone and a fast Fourier transform performed on the signal. Different peaks clearly appear in the frequency spectrum, corresponding to the bowl's different vibrational modes.
 Fig.~\ref{spectraglass}a indicates the frequency spectrum generated by striking the empty bowl Tibet 4. When the bowl is rubbed with a leather wrapped mallet, the lowest mode is excited along with its harmonics, an effect known as a mode ``lock in'' \cite{Akay:2002}. The frequency spectrum of Tibet 4 when rubbed by a leather mallet is presented in Fig.~\ref{spectraglass}b. Due to the bowl asymmetry, two peaks separated by several Hz arise and a beating mode is heard. This split is highlighted in a magnification of the first peak $f_{(2,0)}$ in the inset of the figure. The deformation shapes are the same with both frequencies but there is horizontal angular shift observed to be $\pi/4$ between them for the fundamental modes $(2,0)$ and $\pi/2n$ for other $(n,0)$ modes.
Finally, we note that, owing to the relative squatness of the bowls and the associated high energetic
penalty of modes with $m \ne 0$, only modes $(n,0)$ were excited; henceforth, such modes are
denoted simply by $n$.

\begin{figure}[h!]
\begin{center}
\includegraphics[width=12cm]{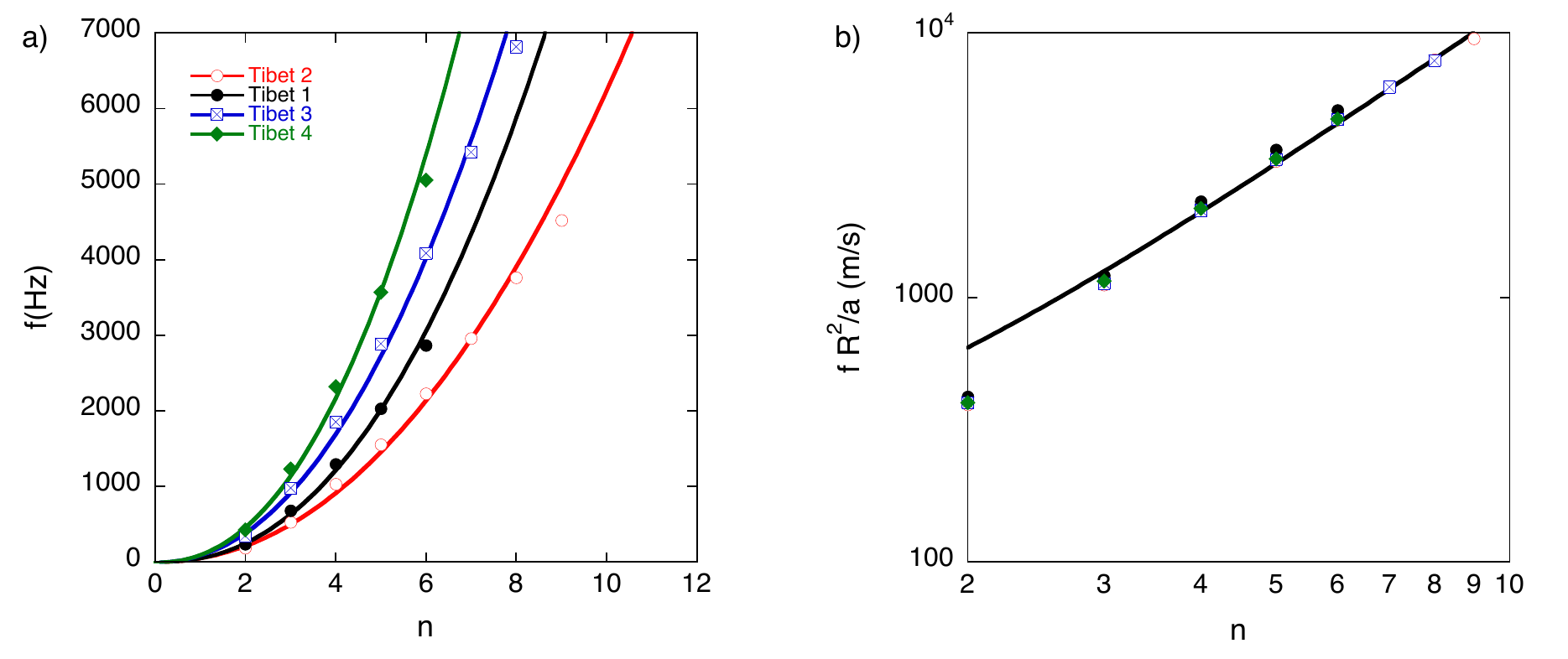}
\end{center}
  \caption{ a) Measurements of the resonant frequencies of the different deformation modes $n$ for the 4 different Tibetan bowls. b)  Characteristic speed $fR^2/a$ as a function of deformation mode $n$ for the 4 different bowls. Data are fit by a power 2 curve indicating that $f_n \propto n^2 a/ R^2$, consistent with Eq. \ref{fScaling}.}
  \label{fvsn}
\end{figure}

We can also find a relation between the different mode frequencies. Since the speed of the bending wave is proportional to the square root of the frequency, $v\propto \sqrt{f}$ and since $\lambda =v/f$, we expect $\lambda \propto 1/\sqrt{fa}$. For the mode $n$, we thus have $2\pi R=n\lambda_n$. The frequency of this mode $n$ should thus be proportional to 
\begin{equation}
f \propto \frac{n^2a}{R^2}. \label{fScaling}
\end{equation}
In Fig.~\ref{fvsn}a, resonant frequencies of the 4 bowls are plotted as a function of their corresponding mode $n$ and a power 2 curve is fit onto each curve. In Fig.~\ref{fvsn}b, we collapse all these curves onto a line by plotting the characteristic speed $f R^2/a$ as a function of the mode $n$, thus validating the proposed scaling (\ref{fScaling}).

\begin{figure}[h!]
\begin{center}
\includegraphics[width=12cm]{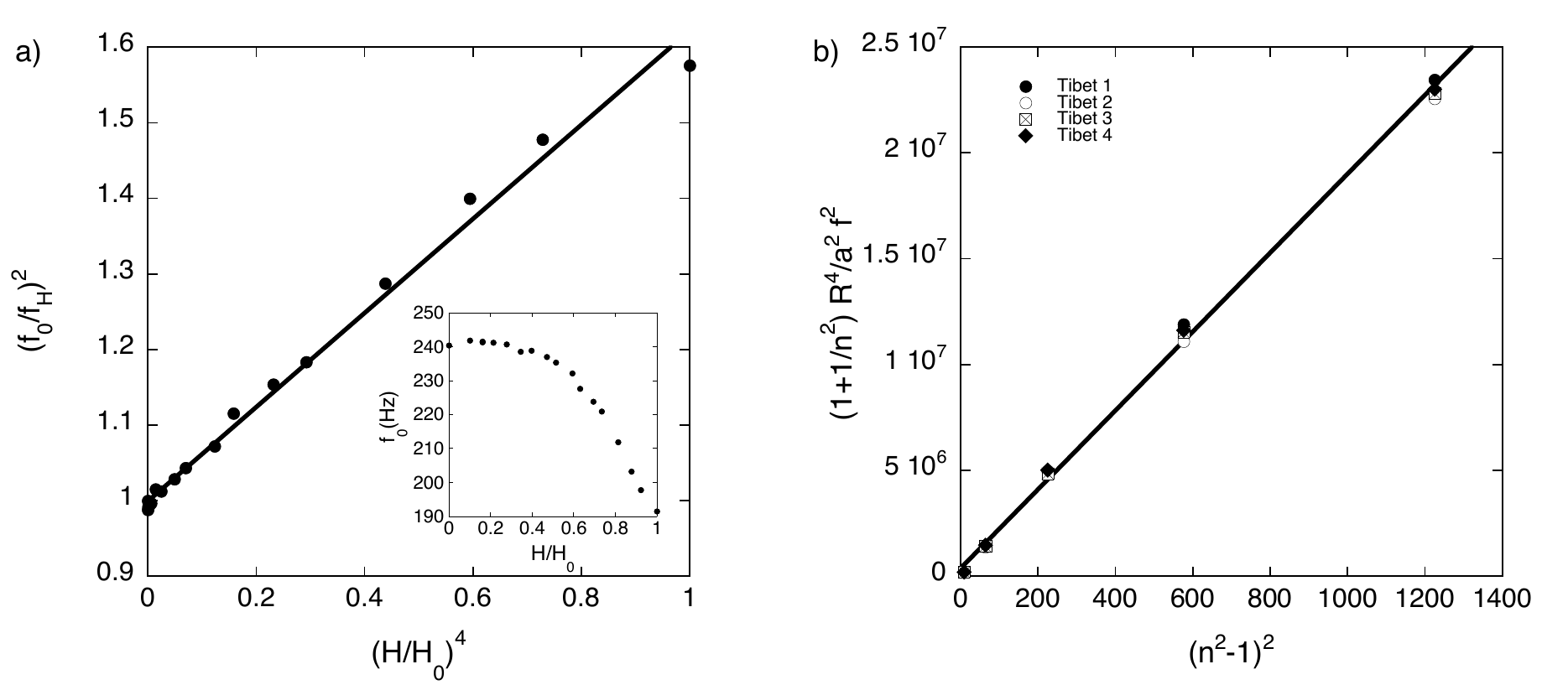}
\end{center}
  \caption{ a) The fundamental frequency of Tibet 1's mode (2,0) as a function of the water height $H/H_0$. According to Eq. \ref{f0surfH} , the slope is equal to  $\frac{\alpha}{5}\frac{\rho_l R}{\rho_s a}$. b) The dependence of the resonant frequencies $f$ of all the bowls on the deformation mode $n$ may be used to infer the ratio $Y/\rho_s$. According to  (\ref{fnm}),  the slope of the line shown is $\frac{1}{48\pi^2}\frac{Y}{\rho_s}$, from which we infer $Y/\rho_s=8.65 \times 10^6\,\rm{Pa.m^3/kg}$.}
  \label{fvshliqYoungDet}
\end{figure}

It is readily observed that the resonant frequencies decrease when liquid is poured into a vessel. In the inset in Fig.~\ref{fvshliqYoungDet}a, we report the measurements of the fundamental frequency of the bowl Tibet 1 as a function of the dimensionless liquid height $H/H_0$. In Fig. \ref{fvshliqYoungDet}a, we report $\left(f_0/f_H\right)^2$ as a function of $(H/H_0)^4$. According to (\ref{f0surfH}), the slope of this curve gives the ratio $\frac{\alpha}{5}\frac{\rho_l R}{\rho_s a}$. In Fig. \ref{fvshliqYoungDet}b, we present the dependence of $(1+1/n^2) (R^2/a)^2 f^2$ on $(n^2-1)^2$ for the deformation modes $n=$2 through 6 of the bowls Tibet 1, 2, 3 and 4. According to (\ref{fnm}), the slope should be equal to $\frac{1}{48\pi^2}\frac{Y}{\rho_s}$. For each value of the abscissa there are 4 measurements corresponding to the 4 Tibetan bowls. Data points from the different Tibetan bowls overlie each other, especially at low $n$, indicating that all bowls have the same value of ratio $Y/\rho_s$, and so are likely made of the same material. A linear fit gives $Y/\rho_s=8.65 \times 10^6\,\rm{Pa.m^3/kg}$.

We can simply estimate the density for each Tibetan bowl by measuring its mass and volume. Masses are measured by a weight scale with an error of 2~g while bowl volumes are deduced by fluid displacement during submersion. The error made on the volume with this method is estimated to be no more than 5$\%$. All such density  measurements are reported in Table \ref{TableTibet}. Taking the mean value of the densities, equal to 8893~kg/m$^3$, we calculate a Young modulus $Y=77 \pm 6\%$~GPa. This value is in the Young's modulus range of glasses and lower than typical brass, copper or bronze alloys for which Y $> 100$~GPa.

\section{Fluid dynamics} \label{SecFluidDynamics}

\subsection{Faraday waves} \label{SecFaradayWaves}

Consider an inviscid fluid of density $\rho$ and surface tension $\sigma$ in a horizontal 
layer of uniform depth $h$
in the presence of a gravitational acceleration $g$. The layer is oscillated vertically in a sinusoidal 
fashion at a forcing frequency $f_0=\omega_0/2\pi$, amplitude $\Delta $ and acceleration $\Gamma g= \Delta \omega^2$. Above a critical forcing acceleration, standing Faraday waves appear on the surface. The associated surface deformation $a(x,y,t)$ can be expressed in terms of the container's eigenmodes $S_m(x,y)$  as
\begin{equation}
a(x,y,t)=\sum_m a_m(t) S_m(x,y),
\end{equation}
with $a_m(t)$ being the oscillating amplitude of the eigenmode $m$. Benjamin and Ursell \cite{Benjamin1954} demonstrate that the coefficients $a_m(t)$ satisfy:
\begin{equation}
\ddot{a}_m+\omega^2_m(1-2\gamma \cos{\omega_0t})a_m=0,  \label{EqMathieu}
\end{equation}
%\begin{equation}
%\ddot{a}_m+\left(g(1-\Gamma \cos{\omega_0t})k_m+\frac{\sigma}{\rho}k_m^3\right)\tanh{(k_mh)}a_m=0
%\end{equation}
where 
\begin{equation}
\omega_m^2=\left(gk_m+\frac{\sigma}{\rho}k_m^3\right)\tanh{(k_mh)} \label{Eqomega}
\end{equation}
represents the classic gravity-capillary wave dispersion relation,
\begin{equation}
\gamma=\frac{\Gamma}{2(1+Bo^{-1})}
\end{equation}
is the dimensionless forcing acceleration, and $Bo=\frac{\rho g}{\sigma k^2}$ is the Bond number. 
$k_m=2\pi/\lambda_m$ is the wave number of the mode $m$, and $\lambda_m$ is the corresponding wavelength. 

In the absence of forcing, $\Gamma=0$, (\ref{EqMathieu}) describes a simple harmonic oscillation with frequency prescribed by (\ref{Eqomega}) corresponding to the free surface vibrations of the liquid. 
When $\Gamma > 0$, (\ref{EqMathieu}) describes a parametric oscillator as the forcing term 
depends on time. The resulting
Faraday waves can be either capillary or gravity waves according to the magnitude of $Bo$; 
specifically, the former and latter arise in the respective limits $Bo \ll1$ and $Bo \gg 1$. Equation \ref{EqMathieu} is known as the Mathieu equation and cannot be solved analytically since one of the terms is time dependent. However, as the forcing is periodic, Floquet theory can be applied to show that an inviscid fluid
layer is always unstable to Faraday waves with 
a frequency $\omega_F$ that is half the forcing frequency  $\omega_0=2\omega_F$  \cite{Benjamin1954}. In the deep water ($kh \gg 1$), capillary wave ($Bo \ll 1)$ limit, we can infer from (\ref{Eqomega}) a Faraday wavelength :
\begin{equation}
\lambda_F=(2\pi)^{1/3}(\sigma/\rho)^{1/3}(f_0/2)^{-2/3}. \label{lambdaF}
\end{equation}

To incorporate the influence of the fluid viscosity, one can add to (\ref{EqMathieu}) a phenomenological dissipation term proportional to the velocity \cite{Kumar1994}:
\begin{equation}
\ddot{a}_m+2\beta \dot{a}_m+\omega^2_m(1-2\gamma \cos{\omega_0t})a_m=0,
\end{equation}
where $\beta $ is the dissipation rate. This dissipation term leads to an acceleration threshold for the Faraday instability. Assuming capillary waves in an unbounded and infinite depth liquid, the critical acceleration needed to induce parametric instability is given by 
\begin{equation}
\Gamma_F \propto \frac{1}{g}(\rho/\sigma)^{1/3}\nu\omega_0^{5/3}, \label{FaradayOnset}
\end{equation}
where $\nu$ is the kinematic viscosity of the fluid \cite{Edwards1994, Douady1990}.

As the forcing amplitude is further increased, the Faraday wave amplitude increases progressively until the waves become chaotic. Ultimately, the waves break and droplets are ejected from the surface. Since drops will be ejected by the breaking of Faraday waves, we expect their diameter to scale as
\begin{equation}
d_m \sim \lambda_F  \sim (\sigma/\rho)^{1/3}f_0^{-2/3} \label{EqDropSize}
\end{equation}
in the capillary wave limit.
Droplet ejection arises when the destabilizing inertial driving force $m \Gamma g$ (with $m\sim \rho \lambda_F^3$) exceeds the stabilizing surface tension force $\pi \lambda_F\sigma$. This implies, via (\ref{EqDropSize}), a threshold acceleration that scales as:
\begin{equation}
\Gamma_d \sim \frac{1}{g} (\sigma/\rho)^{1/3} f_0^{4/3} \label{EqDropEject}
\end{equation}
The range of validity of these scalings will be investigated in our experimental study.

\subsection{Experimental technique} \label{SecExperimentalTechnique}

\begin{figure}[h!]
\begin{center}
\includegraphics[width=10cm]{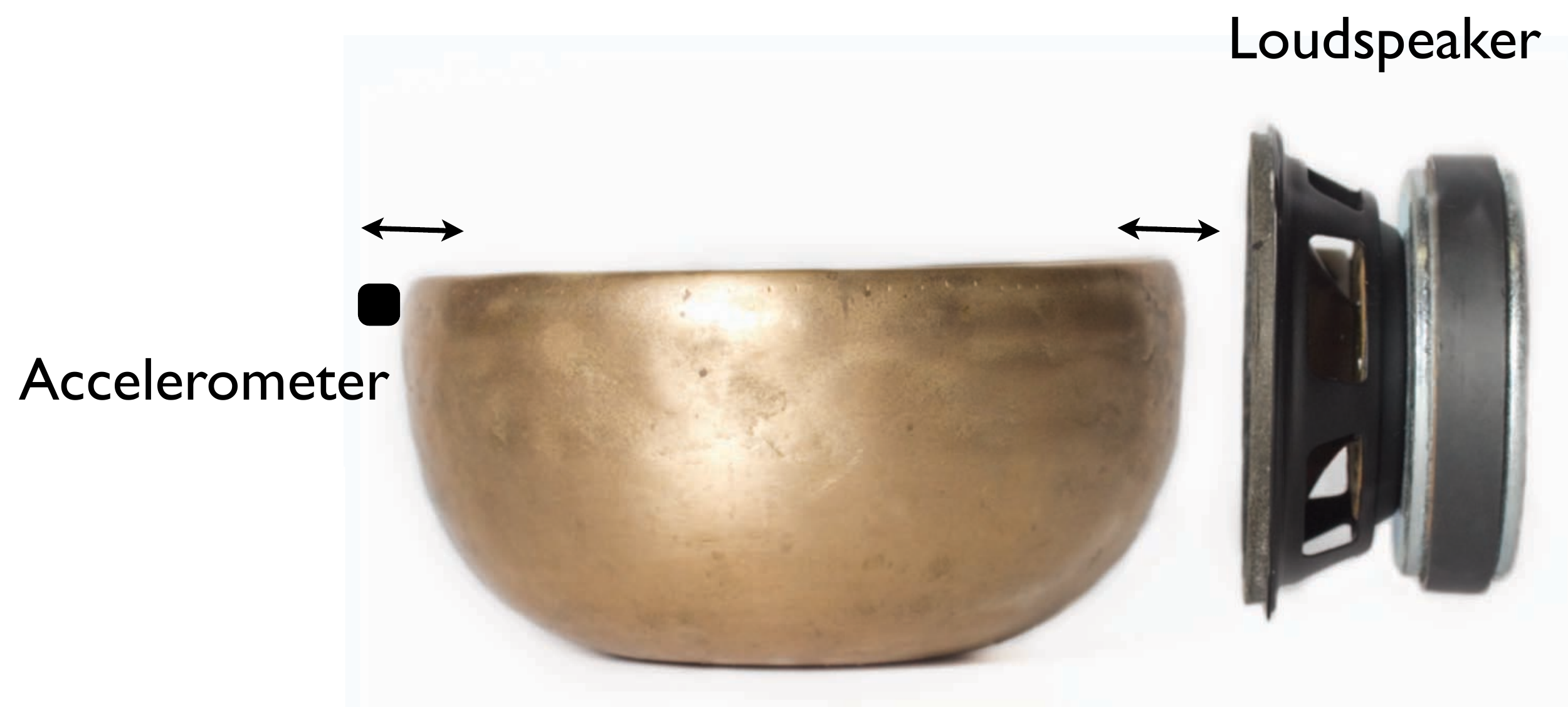}
\end{center}
\caption{Experimental setup. The bowl deformation is excited by a loudspeaker emitting sound at a frequency corresponding to a particular vibrational mode of the bowl. The deformation of the bowl is measured by an accelerometer glued to the bowl's outer wall at the height of the liquid surface. \label{Expapparatus}}
\end{figure}

The experimental setup used for studying the surface waves generated within the Tibetan bowls is presented in Fig.~\ref{Expapparatus}. A loudspeaker is set in front of the bowl, its signal received from a signal function generator then amplified. When the applied signal frequency is close to one of the bowl's resonant frequencies, it oscillates in the corresponding deformation mode. Conveniently, with this method, we can select a single  deformation mode. Recall that by striking or rubbing the bowls, we excited several modes (Fig.~\ref{spectraglass}): moreover, rubbing induced a rotational motion that followed the mallet. We can now examine the vibration-induced flows in a controlled fashion.

In order to extend the range of natural frequencies, we consider Tibetan bowls, wineglasses and soda cans 
(with their tops removed), whose resonant frequencies span a broad range from 50 to 750 Hz. For each, we can vary the resonant frequency by filling it with liquid. The sound emitted by the loudspeaker was not powerful enough to induce signifiant oscillations of the soda can rim, a problem we eliminated by directly connecting the vibrating membrane of the loudspeaker to the rim of the soda can with a rigid rod.

We measure the acceleration of the rim at an antinode by gluing a lightweight accelerometer (PCB-Piezotronics, 352C65) on the bowl's rim at the level of the liquid surface. In the following, we characterize the sinusoidal rim oscillation by the dimensionless acceleration $\Gamma$ defined as the maximal acceleration of the rim normalized by the gravitational acceleration $\Gamma=\Delta \omega^2/g$. For wineglasses, the accelerometer cannot be used since it dramatically alters the resonant frequency. We thus used a light weight strain gauge, whose effect on the resonance frequency is negligible.

The strain gauge system provides a measurement of the local extension length of the rim at an antinode. The length variation of the strain gauge is deduced by measuring its electrical resistance with a Wheatstone bridge. To deduce the acceleration of the radial rim movement, we deduce a relation between the longitudinal extension $\epsilon$ and the radial amplitude of the rim $\Delta$. Then, the acceleration $\Gamma$ can be readily deduced. The validity of this indirect method was tested on Tibet 1. An accelerometer was glued next to a strain gauge at an antinode of the bowl Tibet 1. From the strain gauge measurement, we deduced an acceleration that we compared to the direct measurement from the accelerometer. These two independent measurements match well for a wide range of accelerations.

French \cite{French1982} gives a relation for the convex change of length $\delta l$ of a curved segment of thickness $a$ as a function of the initial mean radius of curvature $R$ and the deformed radius of curvature $r_c$~: $\delta_l=l_0\frac{a}{2}\left(\frac{1}{r_c}-\frac{1}{R} \right)$ where $l_0$ is the initial segment length. Since the deformation is small, we can approximate: $\frac{1}{r_c}\sim\frac{1}{R}+\frac{3x}{R^2}$ where $x$ is the radial displacement of the wall \cite{French1982}. The strain gauge gives the measurement of the convex longitudinal extension of the rim $\epsilon=\delta l/l_0$. We can thus deduce the maximal radial amplitude of the rim deformation \cite{French1982}: $x_m=\frac{2}{3}\frac{R^2}{a}\epsilon$.

\subsection{Surface waves} \label{SecSurfaceWaves}

Two different liquids were used in our experiment, pure water with density $\rho_w=1000$~kg/m$^3$, viscosity $\nu=1$~cSt and surface tension $\sigma=72$~mN/m, and Dow Corning silicone oil with $\rho_o=820$~kg/m$^3$, $\nu=1$~cSt and $\sigma=17.4$~mN/m. The fluid depth and resulting natural deformation frequencies of the bowl were varied. Specific deformation modes of the bowls were excited acoustically. We proceed by reporting the form of the flow induced, specifically, the evolution of the free surface with increasing rim forcing. 

\begin{figure}[h!]
\begin{center}
\includegraphics[width=12cm]{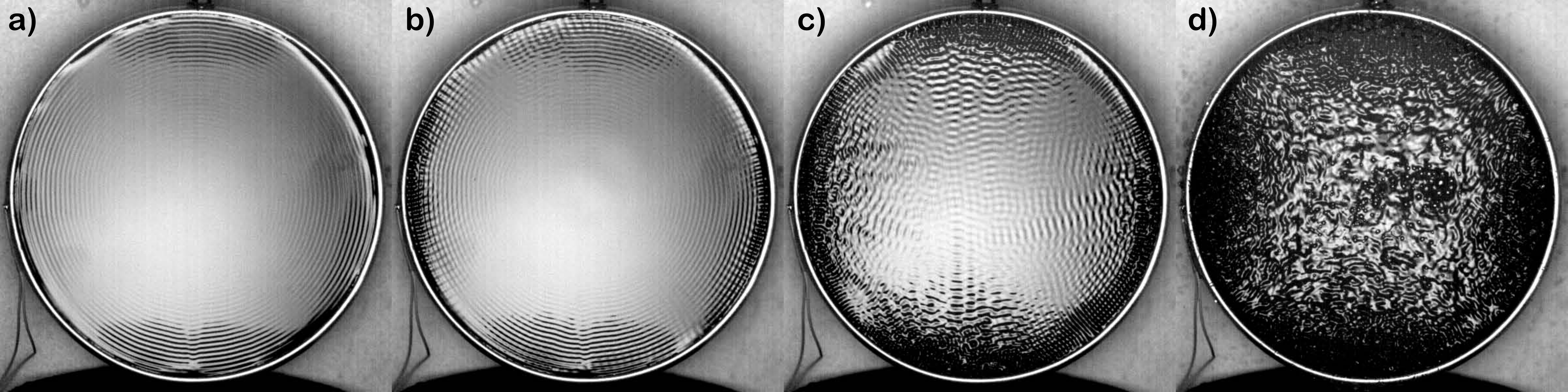}
\end{center}
\caption{Evolution of the surface waves in Tibet 1 bowl filled with water and excited with a frequency $f=188\,\rm{Hz}$ corresponding to its fundamental mode (2,0). The amplitude of deformation is increasing from left to right: a) $\Gamma=1.8$, b) $\Gamma=2.8$, c) $\Gamma=6.2$, d) $\Gamma=16.2. $ \label{T1surf}}
\end{figure}

In Fig.~\ref{T1surf}, we present snapshots of the bowl Tibet 1 resonating in its fundamental deformation mode with different $\Gamma$ when it is completely filled with water. The loudspeaker produces a sinusoidal sound at a frequency $f_0=188 \,\rm{Hz}$ that corresponds to the mode (2,0) with four associated nodes and antinodes. The vibration of the water surface is forced by the horizontal oscillation of the rim. When the amplitude of the rim oscillation is small, axisymmetric progressive capillary waves with frequency commensurate with the excitation frequency appear on the liquid surface. Though almost invisible to the naked eye, they can be readily detected by appropriate lighting of the liquid surface (Fig.~\ref{T1surf}a). When $\Gamma$ is further increased, relatively large amplitude circumferential standing waves appear at the water's edge (Fig.~\ref{T1surf}b). %The amplitude of these waves grows rapidly, and is larger than that of the axial waves; moreover, 
These standing ripples, aligned perpendicular to the bowl's edge, are spaced by approximately a Faraday wavelength $\lambda_F$, as defined in (\ref{lambdaF}). Moreover, their frequency is half that of the axial waves, indicating that these waves correspond to classic cross waves or, equivalently, edge-induced Faraday waves \cite{Faraday1831}.  More complicated wave modes appear at higher excitation amplitude (Fig.~\ref{T1surf}c). At sufficiently high $\Gamma$, the Faraday waves break, and water droplets are ejected from the edge of the vessel (Fig.~\ref{T1surf}d), specifically from the anti-nodes of the oscillating wall. The ejected droplets may bounce, slide, and roll on the water surface before eventually coalescing.

\begin{figure}[h!]
\begin{center}
\includegraphics[width=12cm]{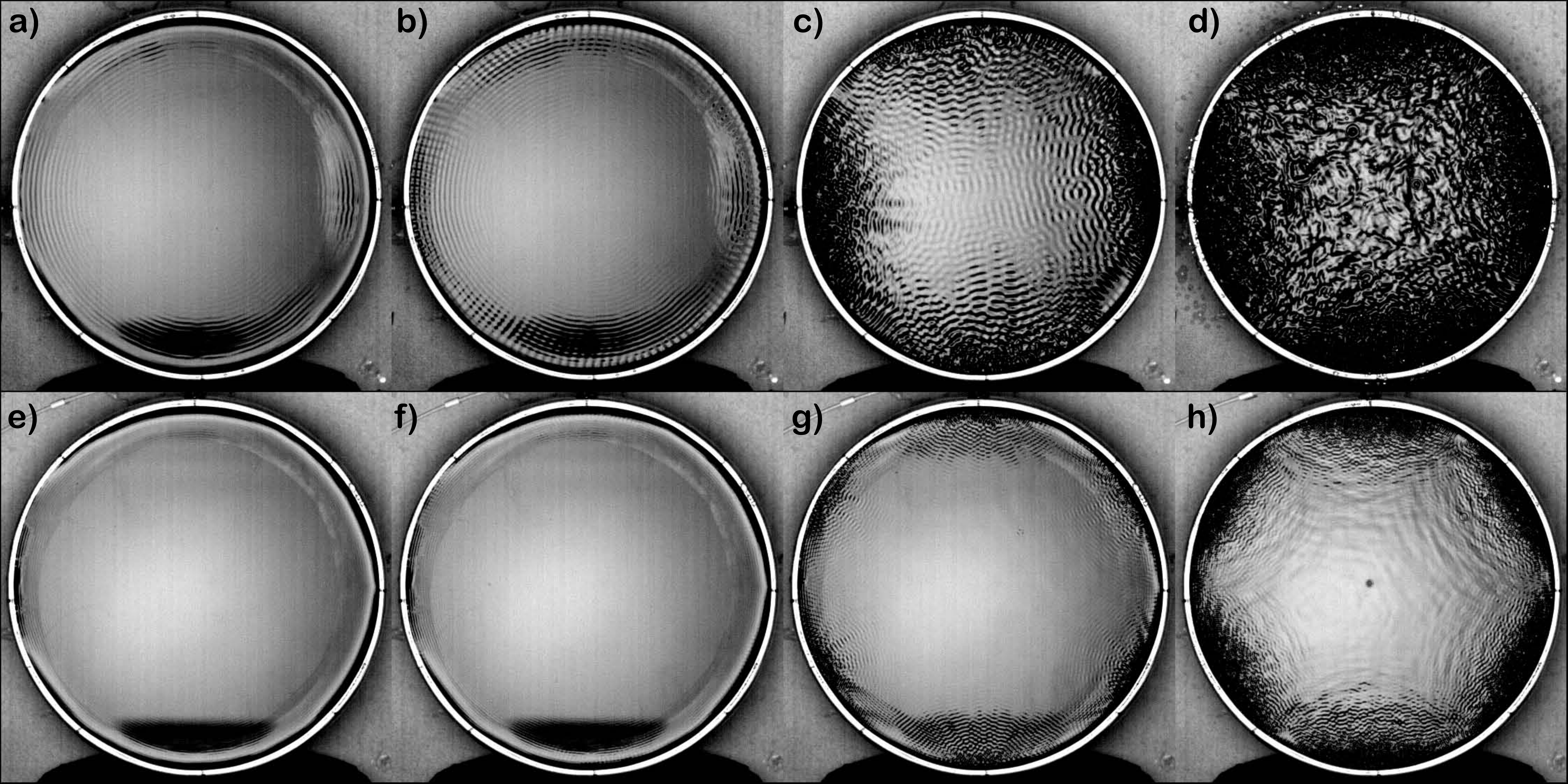}
\end{center}
\caption{Evolution of the surface waves in Tibet 2 bowl filled with water. a-d) Mode (2,0) is excited with a frequency $f=144\,\rm{Hz}$. e-h) Mode (3,0) is excited with $f=524\,\rm{Hz}$. The amplitude of deformation is increasing from left to right. For the deformation mode $n=2$, a) $\Gamma=1.0$, b) $\Gamma=1.7$, c) $\Gamma=5.0$, d) $\Gamma=12.7$. For the deformation mode $n=3$, e) $\Gamma=5.7$, f) $\Gamma=7.7$, g) $\Gamma=15.1$, h) $\Gamma=33.8$. \label{T2surf}}
\end{figure}

One of our bowls (Tibet 2) resonates readily in both modes (2,0) and (3,0). When completely filled with water, the resonant frequencies of its (2,0) and (3,0) modes  are $f=144\,\rm{Hz}$ and $f=524\,\rm{Hz}$, respectively. In Fig.~\ref{T2surf}, we observe the progression of the surface waves with increasing amplitude for each of these modes. Note that for the mode (3,0), since the frequency is higher, the wavelengths are shorter. Moreover, the sound amplitude needed to produce surface waves is necessarily higher for mode (3,0) than (2,0).

The transition from axisymmetric capillary waves to Faraday waves arises at a critical acceleration $\Gamma_F$ readily measured by the accelerometer. This threshold was measured as a function of the forcing frequency, the latter having been tuned to excite the fundamental deformation modes of the bowls with different liquid levels. We also measured this acceleration threshold for the deformation mode (3,0) of the bowl Tibet 2. Higher frequencies were explored with three different wineglasses filled to different levels using the strain gauge system. All the measurements with silicone oil of viscosity $1\,\rm{cSt}$ are presented in Fig.~\ref{G1G2ThreshVSf} (lower curve). Consistent with (\ref{FaradayOnset}), the data suggest a dependence $\Gamma_F \propto f^{5/3}$. In Fig.~\ref{Gamma1-2_all}a, we report our measurements of $\Gamma_F$ as a function of frequency for both silicone oil and distilled water. Each data set is fit by a $5/3$ power law. Prefactors of $3.5 \times 10^{-4}$ for water and $1.7 \times 10^{-4}$ for 1 cSt silicone oil were inferred.

\subsection{Surface fracture} \label{SecSurfaceFracture}

\begin{figure}[h!]
\begin{center}
\includegraphics[width=9cm]{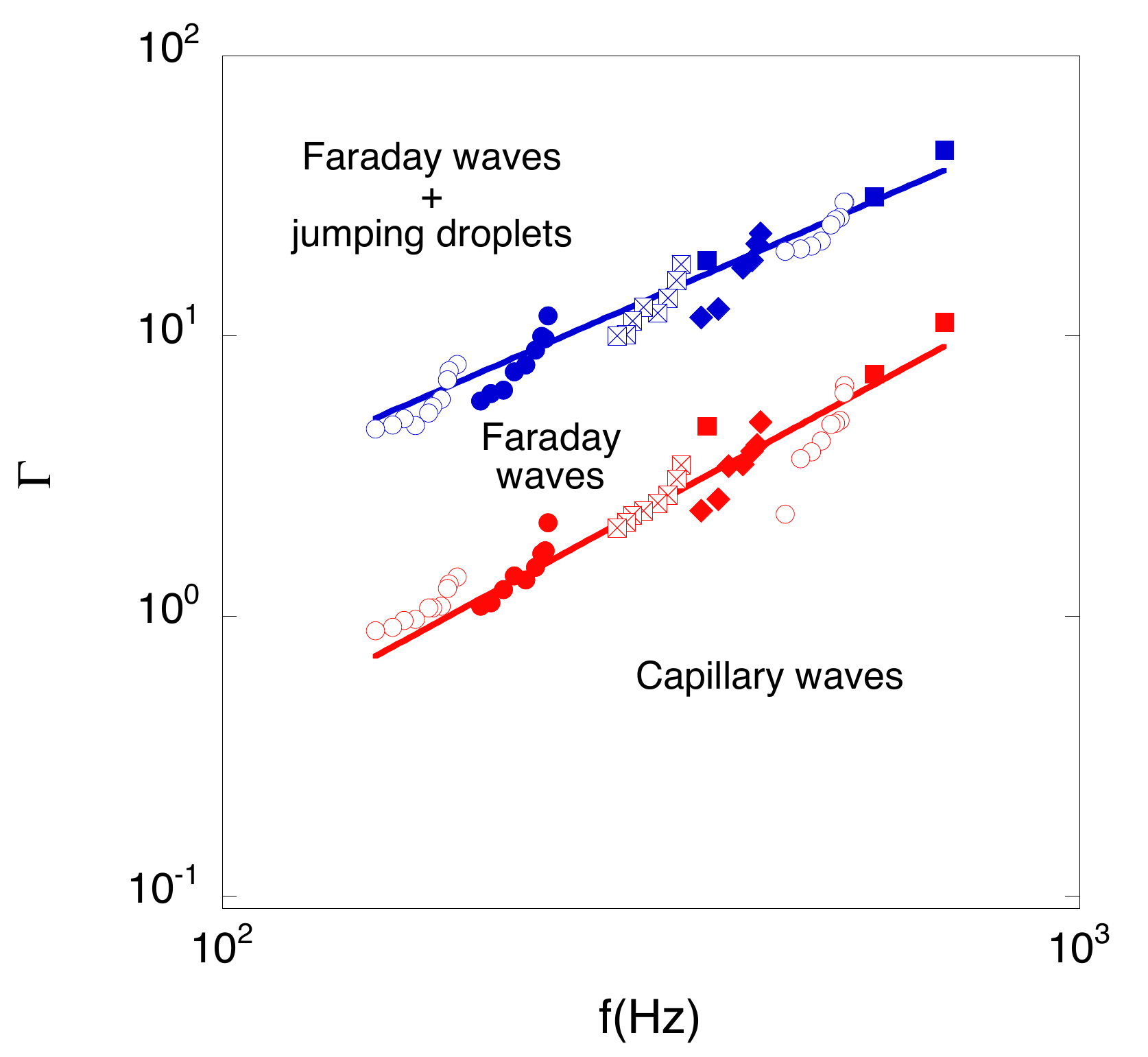}
\end{center}
\caption{Phase diagram indicating the critical acceleration $\Gamma_F=\Delta \omega^2/g$ above which axial capillary waves give way to Faraday waves, and $\Gamma_d$ above which the Faraday waves break. The $\bullet$,  $\circ$, $\boxtimes$ and $\blacklozenge$  symbols correspond, respectively, to the measurements made with the bowls Tibet 1, 2, 3 4 and the  $\blacksquare$ to wineglasses filled with different levels of $1\,\rm{cSt}$ silicone oil. The first acceleration threshold (lower curve) scales as $\Gamma_F\propto f^{5/3}$ and the second (upper curve) as $\Gamma_d\propto f^{4/3}$. \label{G1G2ThreshVSf}}
\end{figure}

When the Faraday waves become sufficiently large, they break, leading to droplet ejection.  A second critical acceleration can thus be measured, $\Gamma_d$, above which droplets are ejected from the surface. The droplet ejection starts with very few droplets ejected, then the ejection rate increases with forcing amplitude. Our criterion for onset is that at least two droplets are ejected in a 15 second time interval. The dependence of 
$\Gamma_d$ on $f$ is presented in Fig.~\ref{G1G2ThreshVSf} (upper curve) for bowls and wine glasses 
filled with different
levels of 1~cSt silicone oil. The drop ejection threshold scales as $\Gamma_d \propto f^{4/3}$, which is consistent with the scaling presented in (\ref{EqDropEject}).

\begin{figure}[h!]
\begin{center}
\includegraphics[width=12cm]{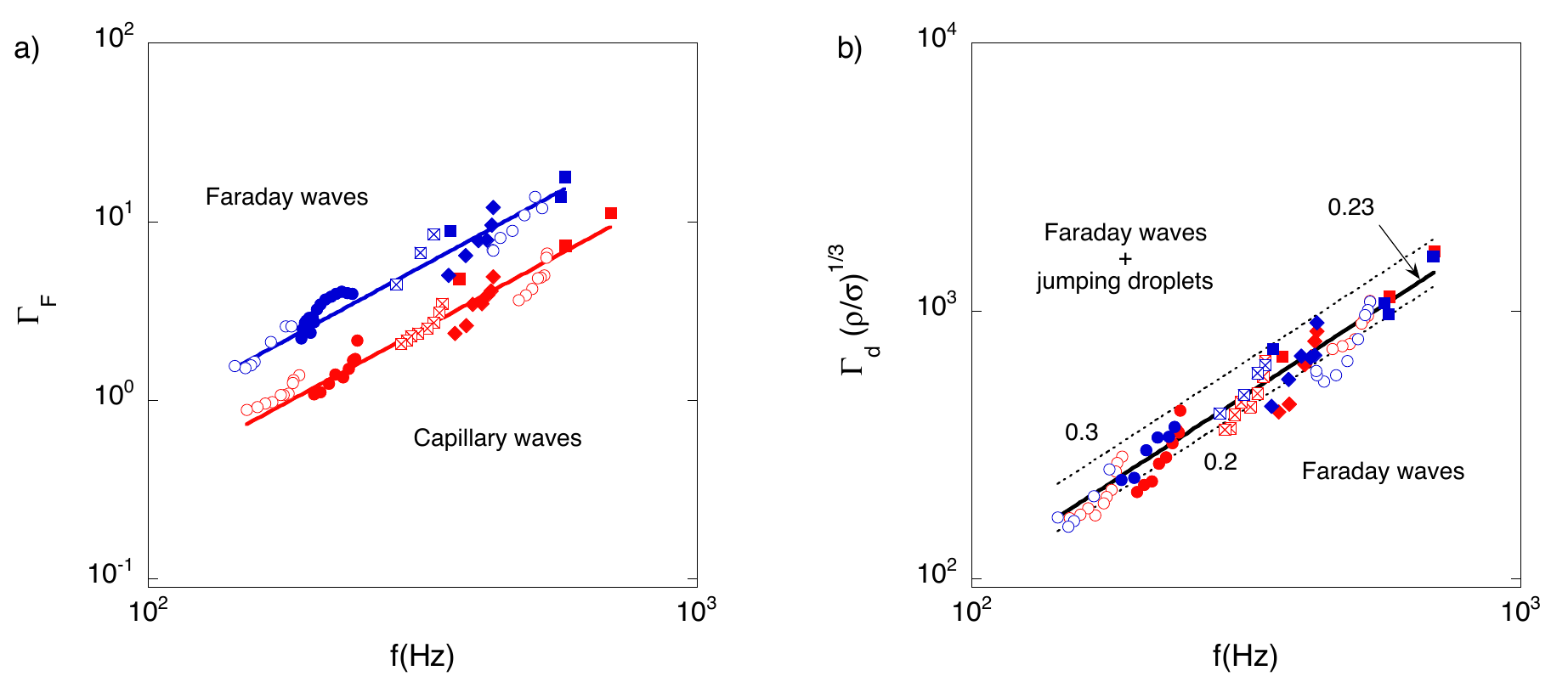}
\end{center}
\caption{Dependence of acceleration thresholds on the frequency $f$ for our vessels filled with different levels of pure water (blue symbols) and silicone oil of viscosity  $\nu=1\,\rm{cSt}$ (red symbols). The $\bullet$,  $\circ$, $\boxtimes$ and $\blacklozenge$  symbols correspond, respectively, to the measurements made with the bowls Tibet 1, 2, 3 4 and the  $\blacksquare$ to wineglasses. a) Each data set of Faraday threshold measurements, $\Gamma_F$, is fit by a 5/3 power law consistent with (\ref{FaradayOnset}). b) The droplet ejection threshold, $\Gamma_d$, is consistent with the scaling (\ref{EqDropEject}). \label{Gamma1-2_all}}
\end{figure}

In Fig.~\ref{Gamma1-2_all}b, our measurements of the dependence of $\Gamma_d$ on $f$ are reported for 
bowls filled with different levels of either 1 cSt silicone oil or water. The two $\Gamma_d$ curves collapse when we use the scaling law (\ref{EqDropEject}) expected to apply for vertical forcing. Specifically, we find
\begin{equation}
\Gamma_d\sim0.23(\sigma/\rho)^{1/3} f^{4/3}. \label{OurEqDropEject}
\end{equation}
The droplet ejection acceleration threshold is thus in accord with measurements of Goodridge \cite{Goodridge1997} and Puthenveettil \cite{Puthenveettil2009}, even though our forcing is horizontal rather than vertical. Moreover, our prefactor is consistent with the results of both, who reported values between 0.2 and 0.3.

\begin{figure}[h!]
\begin{center}
\includegraphics[width=12cm]{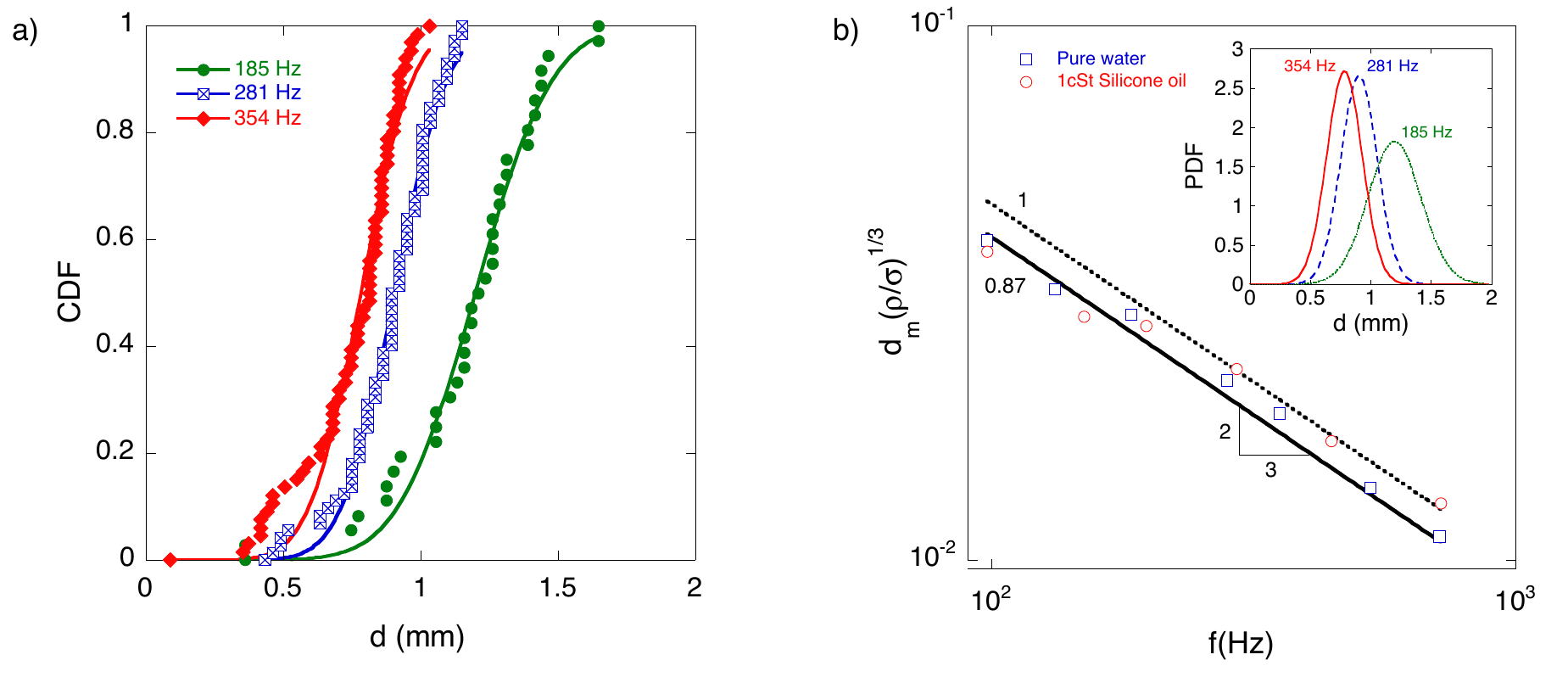}
\end{center}
\caption{a) Three cumulative distribution functions of ejected droplet sizes for the bowls Tibet 1, 3 and 4, fully filled with pure water and resonating in their fundamental deformation modes (2,0). The lines are the curves fit by an equation of a cumulative distribution function for the Gaussian distribution.  b) Mean size of the ejected droplets as a function of the frequency of the oscillating vessel, either a Tibetan bowl, wineglass or soda can. The two leftmost data points are from the measurements made with the soda cans, the two rightmost from the wineglass. Error bars are the size of our symbols. The black curve indicates a power law dependence with slope -2/3 and prefactor 0.87. The dotted line has the same slope, but a prefactor of 1. Inset: The corresponding Gaussian functions inferred from the cumulative distribution functions. \label{dropsize}}
\end{figure}

The diameter of the ejected droplets was measured for several forcing frequencies. The Tibetan bowls were fully filled with liquid, either 1 cSt silicone oil or pure water. Resonant deformation modes were then excited by the loudspeaker emitting sinusoidal signals at the appropriate resonant frequency. The level of sound was adjusted so that the acceleration of the rim at the antinodes of the bowl were just above the threshold for droplet ejection, $\Gamma_d$. A high speed video camera  (Phantom) was used to record the ejected droplets, from which drop size measurements were taken. We used a liquid-filled glass in order to extend the frequency range to 720 Hz. Soda cans have very thin walls, and very low resonant frequencies. We were thus able to measure droplet sizes for the two liquids at a frequency of 98 Hz. Three cumulative distributions of ejected droplet sizes are presented on Fig.~\ref{dropsize}a for the bowls Tibet 1, 3 and 4 resonating in their fundamental deformation modes (2,0). Assuming that these distributions are Gaussian, appropriate fits to the cumulative distribution functions yield the parameters of the Gaussian distribution functions plotted in the inset of Fig.~\ref{dropsize}b. The dependencies of the mean droplet size on the forcing frequency for the two liquids are presented in Fig.~\ref{dropsize}b. Equation (\ref{EqDropSize}) adequately collapses our data, provided we choose a prefactor of $0.87$ . By way of comparison, Puthenveettil found 0.92 for their experiments with water, and 1.01 with Perfluoro-compound FC-72 liquid. Donnelly \textit{et al.} \cite{Donnelly2004} found a prefactor of 0.98 
from his measurements of aerosol water droplets.

\subsection{Bouncing droplets} \label{SecBouncingDroplets}

With a more viscous fluid (e.g. 10~cSt silicone oil), the waves are less pronounced, and the fluid is more
strongly coupled to the vibrating sidewalls; specifically, more of the surface oscillates up and down near the wall's anti-node. When a droplet of the same liquid is deposited on the surface, it may bounce, levitated by the underlying wave field. Such sustained levitation was not observed in the Tibetan singing bowl with liquid viscosities lower than 10 cSt.

\begin{figure}[h!]
\begin{center}
\includegraphics[width=12cm]{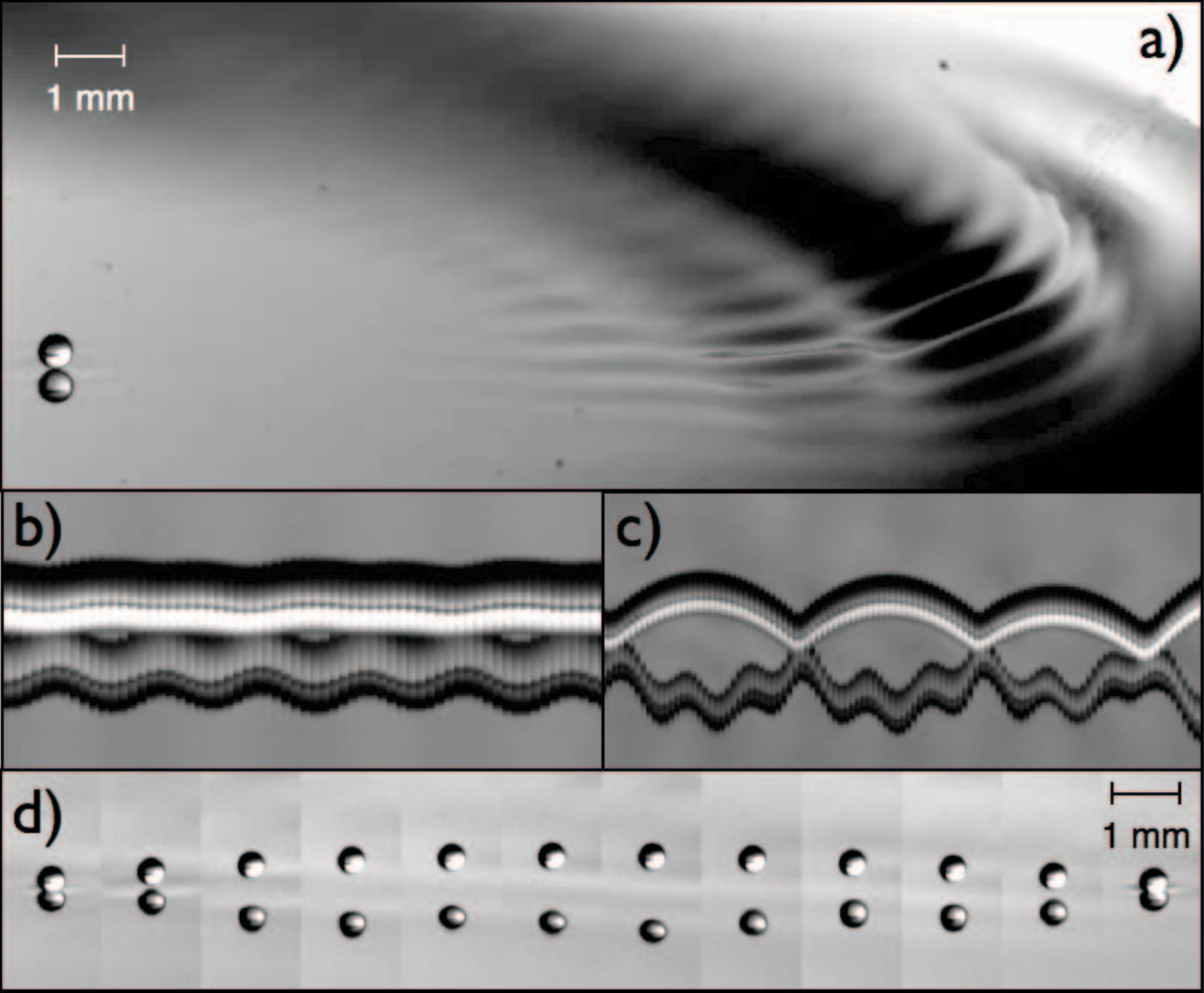}
\end{center}
\caption{a) Still image of a droplet of 0.5 mm diameter bouncing on a surface oscillating at 188 Hz, the frequency of the rim. Both droplet and bath are 10 cSt Silicon oil. Spatio-temporal diagrams indicate the vertical trajectory of droplets of diameter b) 0.5 mm and c) 0.35 mm. Time elapses from left to right and the drop's reflection is apparent. d) A single bounce of a droplet of diameter 0.35 mm is illustrated by an image sequence. Images are spaced by 1 ms.\label{bouncingdrop}}
\end{figure}

In Fig.~\ref{bouncingdrop}a, we present a still image of a drop of diameter 0.5 mm bouncing on the liquid surface inside the bowl (Tibet 1) resonating at a frequency of $f_0=188\,\rm{Hz}$. The drop has been made by dipping then extracting a syringe needle, on the tip of which a capillary bridge forms and breaks, leaving a drop that bounces near the oscillating rim. A movie of the bouncing was recorded, then vertical slices of each image through the droplet centerline juxtaposed. We can thus construct an image illustrating the dynamics of the droplet (Fig.~\ref{bouncingdrop}b). In this case, the drop experience two bounces of slightly different amplitude while the liquid surface (and the rim) oscillates twice. When a smaller droplet (of diameter 0.35~mm) is placed on the oscillating surface, the bouncing motion can be more complex. In Fig.~\ref{bouncingdrop}c, we see that the droplet bounces only once during three oscillations of the surface. Fig.~\ref{bouncingdrop}d illustrates the corresponding trajectory. We note that the bouncing motion of sufficiently small drops can become chaotic.

We sought to sustain walking droplets in our system. Once the horizontal forcing amplitude $\Gamma$ is just above the Faraday wave threshold $\Gamma_F$, circumferential Faraday waves are sustained at the edge of the vessel. Then, when $\Gamma$ is increased, Faraday waves propagate progressively towards the center of the vessel, their amplitude damped by viscosity. Beyond these waves, the liquid surface was quiescent unless perturbed by a bouncing droplet, in which case it could sustain a field of Faraday waves. This phenomenon was observed in the bowl Tibet 2 almost fully filled with 10 cSt silicone oil ($f_0$=140 Hz). Just beyond the Faraday waves, a bouncing droplet of diameter 500 $\mu$m was made such that its bouncing frequency corresponded to the Faraday wave frequency, that is, half the frequency of the vibrating rim. Such droplets were unable to excite sufficiently large Faraday waves to enable them to walk. We note that the usual range of walker diameters is between 650 $\mu$m and 850 $\mu$m \cite{protiere:2005, eddi:2008}: their mass is thus 4 times larger than that of our drops.

\section{Conclusion} \label{SecConclusion}

We have presented the results of an experimental investigation of the Tibetan singing bowl, its acoustics and hydrodynamics. Its acoustical properties are similar to those of a wine glass, but its relatively low vibration frequency makes it a more efficient generator of edge-induced Faraday waves and droplet generation via surface fracture.

Our observations of the bowl acoustics have been rationalized by adapting French's \cite{French1982} theory of the singing wine glass. This model allowed us to characterize the bowl acoustics and infer the Young modulus of the alloy constituting our antique bowls. The value we found, $Y=77 \pm 6\%$~GPa is in the range of glass, somewhat smaller than typical brass, copper or bronze alloys. This low value of $Y$ and associated low resonant frequency is a critical component in the hydrodynamic behavior of singing bowls: bowls with high fundamental frequencies are, like the wine glass, relatively inefficient generators of droplets.

Particular attention has been given to the Faraday waves produced when a critical acceleration of the horizontal rim oscillation is exceeded. These have been shown to be due to a destabilization of the axial capillary waves similar to those observed and studied theoretically \cite{Hsieh2000, Garrett:1970}. The acceleration threshold for droplet ejection has also been investigated and rationalized by simple scaling arguments. Droplet size was shown to be proportional to the Faraday wavelength, and our measurements were consistent with those on a vertically shaken liquid surface \cite{Puthenveettil2009, Goodridge1997, Donnelly2004}.

We have demonstrated that, following their creation via surface fracture, droplets may skip across or roll along the surface of fluid contained within a singing bowl. Moreover, careful choice of fluid properties and 
droplet position introduces the possibility of stable bouncing states reminiscent of those on a vertically driven free surface \cite{Couder2005}. However, stable walking droplets and their concomitant quantum behaviour 
were not observed. Nevertheless, in developing hydrodynamic analogues of quantum systems, the edge-forcing examined here may be valuable in presenting a lateral gradient in proximity to Faraday threshold.

\vspace*{0.3in}

\noindent{{\bf Acknowledgements}}

\vspace*{0.1in}

The authors thank Rosie Warburton for bringing this problem to our attention, and for supplying the bowls for our study. We are grateful to Jim Bales (of MIT's Edgerton Center), Barbara Hughey and St\'ephane Dorbolo for assistance and fruitful discussions. Denis Terwagne thanks the University of Li\`ege and the GRASP for financial support. This research was supported by the National Science Foundation through grant 
CBET-0966452.

%Bouncing droplets could be sustained on the liquid surface of the bowl vibrating. For a given rim oscillation, the bouncing periodicity depends on the size of the droplet. As the liquid surface is subject to an horizontal acceleration that decreases towards the center, the Faraday instability is initiated at the rim. It would be worth studying bouncing droplets on such a surface produced by a wave maker for various liquid viscosities. Large bouncing droplet produces a wave field of large amplitude which symmetry could be brake by the forcing gradient that could influence the bouncing motion of the droplet. Moreover, It would be of interest to characterize the liquid surface with an optical method for surface reconstruction \cite{Moisy:2009} to fully understand the surface motion induced by the wave maker.

%\clearpage

%\clearpage
%\bibliographystyle{DEAc}
%\bibliography{biblio}

\end{document}